\documentclass[12pt]{article}

\newcommand{\be}{\begin{equation}}
\newcommand{\ee}{\end{equation}}
\newcommand{\bdm}{\begin{displaymath}}
\newcommand{\edm}{\end{displaymath}}
\newcommand{\bean}{\begin{eqnarray}}
\newcommand{\eean}{\end{eqnarray}}
\newcommand{\dd}{\partial}
\newcommand{\ra}{\rightarrow}

\newcommand{\vf}{\varphi}

\def\p{\partial}

\def\half{\frac{1}{2}}

\def\bea{\begin{eqnarray}}
\def\eea{\end{eqnarray}}

\begin{document}

\title{\bf{Dimensional Reduction of Gravity and Relation between
Static States, \\
 Cosmologies and Waves}}

 \author {  {\bf V. de Alfaro }\thanks{vda@to.infn.it}~ and
{\bf A.T. Filippov} \thanks{filippov@theor.jinr.ru}~\\
{\small {\it $*$ Dip. di Fisica Teorica and INFN, via Giuria 1,
I-10125 Torino }} \\
{\small {\it $*$ Accademia delle Scienze di Torino, via Maria Vittoria 3,
I-10123 Torino}} \\
{\small {\it $+$ Joint Institute for Nuclear Research, Dubna, Moscow
Region, RU-141980 }   }  }

\date{}

\maketitle

\begin{abstract}

We introduce generalized dimensional reductions of an integrable
1+1-dimensio\-nal dilaton gravity coupled to matter down to
  one-dimensional static states (black holes in particular), cosmological
models and waves. An unusual feature of these reductions is the fact that
the wave solutions  depend on two variables -- space and time.  They are
obtained here both by reducing the moduli space (available due to complete
integrability) and by a generalized separation of variables (applicable
also to nonintegrable models and to higher-dimensional theories). Among
  these new wave-like solutions we have found a class of solutions for which
the matter fields are finite everywhere in space-time, including infinity.

These considerations clearly demonstrate that a deep connection exists
between  static states, cosmologies and waves. We argue that it should
exist in realistic higher-dimensional theories as well.
  Among other things we also briefly outline the relations existing between
  the low-dimensional models that we have discussed here
  and the realistic higher-dimensional ones.

This paper develops further some ideas already present in our previous
papers. We briefly reproduce here (without proof) their main results
in a more concise form and give an important generalization.

\end{abstract}

\section{Introduction}
The recent observations of the acceleration of our Universe demonstrated
the need to consider a wider multitude of cosmological models that
possibly can explain the origin of the dark energy\footnote{For a recent
general review of the dark energy problem see \cite{Padmanabhan}. Many
cosmological models attempting to explain the origin of the dark energy
   are reviewed in \cite{Copeland}, \cite{Nojiri}.}.
   Although a model independent analysis
of the present observational data \cite{Sahni} is compatible with the
simplest assumption of supplementing the Einstein gravity by the Einstein
cosmological constant term, this does not clarify the physical or
geometrical origin of the dark energy. Many attempts to understand the
dark energy origin are somehow related to an evaluation of the role of
different scalar matter fields in gravity theory and of the (not
necessarily constant) cosmological potential. They may originate from
dimensional reductions of gravity or supergravity theories motivated by
superstrings (for a general pre-dark-energy-era  reviews of string
motivated cosmological models see, e.g., \cite{Lid}, \cite{Venezia}).

One may also argue that a detailed understanding of nonperturbative
features of even the standard gravity coupled to matter are not so well
understood and should deserve a careful investigation in completely
integrable models of gravity coupled to matter fields. For long time, many
exact solution of gravity not coupled to matter field were found and
studied (see a comprehensive review \cite{Schmutzer}) and, more recently,
 integrable models describing some important aspects of gravity theory were
discovered (see, e.g., \cite{BZ} - \cite{Alekseev} and references
  therein). The first model that incorporated some most important properties
  of black holes coupled to many scalar matter fields \cite{CGHS} appeared
  15 years ago and was rather popular as it allowed to understand some
features of black hole evolution. In cosmology, only one-dimensional
models were studied and their relation to the two-dimensional integrable
models of gravity was not discussed. This may possibly be explained by the
fact that the string-inspired CGHS model \cite{CGHS} is not realistic
  enough, having minimal coupling of scalar fields and
  constant cosmological potential (not corresponding to the cosmological
  constant in higher dimensions). As a result, the geometry of the
two-dimensional space-time is trivial and cosmologies and waves in this
model are not realistic at all. Much more realistic two-dimensional model
were obtained from spherically symmetric gravity coupled to matter but
they are not integrable. Realistic two-dimensional cosmological models
  were also inspired by string theory and some of
  them are related to integrable $\sigma$-models (see, e.g. \cite{Venezia}).
However, they are not too well suited for considering black holes and
other realistic static states since the cosmological potential in them is
  identically zero.

Thus, there exists a serious motivation to study more complex
two-dimensional models of gravity coupled to matter fields and compare
them to more realistic approximations that are able to describe real black
holes, cosmologies and waves. This paper continues our studies of the
dimensionally reduced description of high dimensional branes (black holes
and other static states in particular), cosmological theories, and some
waves coupled to gravity, which was presented in papers \cite{VDA1} -
\cite{VDA3}. The first motive of this paper is to illustrate the different
problems and approaches to their solution by a detailed consideration of
the explicitly integrable $N$-Liouville model of gravity coupled to
matter, which is close to some realistic theories and can give description
of real black holes and cosmologies in one dimension. The second motive is
to demonstrate that the most general approach to dimensional reduction
from dimension (1+1) to (1+0) or (0+1) may be based on the properly
formulated generalized separation of variables. We also will try to
demonstrate more explicitly that a deep connection exists between black
holes (or, more generally, static states of gravity coupled to matter),
various cosmological models, and, possibly, the simplest waves of scalar
matter coupled to gravity.

Some time ago we observed that a naive dimensional reduction from the
(1+1)-dimensional to the one-dimensional dilaton gravity does not produce
the standard (FRW) cosmology. This fact was mentioned in \cite{VDA1},
where we found it necessary to return to the `parent'  higher-dimensional
spherically symmetric gravity (equivalent to the (1+1)-dimensional theory)
and directly reduce it to the homogeneous and isotropic FRW cosmologies.
This looks strange because the FRW cosmologies are spherically symmetric
and the (1+1)-dimensional dilaton gravity describes all possible
spherically symmetric solutions of the `parent' theory. Thus the problem
-- how to obtain the standard cosmology directly from the dilaton gravity
by some dimensional reduction -- remained unsolved. Most misleading is the
fact that the naive reduction produces the correct static state of pure
gravity -- the external part of the Schwarzschild black hole. The naive
cosmology is simply the internal part of the black hole that has nothing
to do with the FRW cosmology. The presence of matter fields significantly
changes the picture -- instead of the black hole there emerges a static
state of matter having no horizon and the corresponding cosmology is one
of the FRW cosmologies (in fact, it is the closed one). This result was
obtained in an old paper \cite{ATF4} but nobody (including the author)
appreciated its paradoxical meaning\footnote{Note that the connection
 between the
 Schwarzschild black hole and some cosmology was first mentioned in
\cite{Snyder}. This connection was implicitly touched upon in the papers
\cite{CAF1}. A more general example demonstrating a close connection
between static spherically symmetric states of scalar matter coupled to
gravity and a spherically symmetric closed cosmology was presented in
\cite{ATF4}, although its meaning was not discussed in detail. At the same
time, similar relations between some static $p$-branes and cosmologies
  were found, see \cite{Lukas} - \cite{Pope}.}.

Attempting to understand this we proposed \cite{ATF1} to obtain the FRW
cosmology by use of a more general dimensional reduction, in which the
metric and dilaton of (effectively one-dimensional)  black holes and
cosmologies depend on two variables. Although in this consideration we
still used  the connection of the two-dimensional metric to the
   higher-dimensional one,
   we concluded that there may exist direct dimensional
reductions from a two-dimensional theory to its one-dimensional
descendants. Another lesson is that there may exist many different
generalized dimensional reductions producing different static and
cosmological solutions that are effectively one-dimensional but formally
depend on two variables.

Note in passing that, to the best of our knowledge, this problem was never
raised in the literature. This may be possibly explained by the fact that,
for many years, the studies of black holes and those of cosmology were
using very different approaches.  The well known fact that the inner part
of the Schwarzschild black hole is some cosmology was considered a
curiosity devoid of serious meaning. While the theory of black holes was
attempting to understand more and more complex theoretical models (from a
spherically symmetric collapsing black hole  to rotating axial ones), the
  basis of the cosmological theory was given by the FRW model and the
mainstream cosmology mostly concentrated on astrophysical and
observational aspects and their adaptation to modern physics. This is
recently changing due to the dark energy problem and
   seems to be able to give a sufficient motivation
   to searches for a common background for all
the three main objects of the modern gravity theory -- static states
(black holes, in particular), cosmological models and gravitational waves
(or, waves of matter coupled to gravity).

In an unpublished paper \cite{VDA2} we tried to develop the above
mentioned ideas in a more systematic way. For example, we envisaged the
idea that it is important to take into account the `surface' (or `total
derivative') terms that appear in the process of dimensional reduction. We
also observed that some standard 'gauge' (`coordinate')  fixings may be
dangerous in this context because they significantly restrict the number
of possible dimensional reductions. However we were far from understanding
the whole problem. For instance, in paper \cite{VDA2} we still considered
only naive dimensional reductions of the integrable $N$-Liouville theory
 `looking at, but not seeing' (so to speak) a new class of one
dimensional solutions.

In \cite{ATF2} it was mentioned that the structure of the solutions of the
integrable $N$-Liouville theory allows for a very simple interpretation of
the dimensional reduction as a reduction in the moduli space. This
immediately shows that in addition to static and cosmological solutions
there exists a class of wave-like solutions. We attempted to carry on a
more detailed investigation of these states in paper \cite{VDA3}, which
was not published mainly because the approach used in that paper seemed to
be applicable to integrable models only.

Here we remake
 paper \cite{VDA3} all over again in the light of the
results of \cite{ATF2} and \cite{ATF3} and relate the `dynamical
dimensional reduction' of integrable models to more general reductions,
based on the generalized separation of variables (which can in principle
be applied to nonintegrable theories as well). We also study in some
detail the wave-like solutions that can be obtained in both approaches and
discuss some nonsingular waves of scalar matter coupled to gravity.

\section{General Relations between two-dimensional and \\
one-di\-men\-sio\-nal Lagrangians}

Let us consider a rather general two-dimensional Lagrangian
 \be
 \label{aa.010}
 {\cal L}^{(2)} \,=\, \sqrt{-g} \, \bigl\{ \varphi\,
R\,+\, V(\varphi, \psi)\, +\,\sum_{m=3}^N \, Z_{mn}(\varphi,\psi)
\,\nabla\psi_m\, \nabla \psi_n\, \bigr\}\,,
 \ee
 where $g_{ij}$ is
a generic (1+1)-dimensional metric with signature (-1,1), $\varphi$ is the
dilaton field, $R$ is the Ricci curvature, $V(\varphi, \psi)$ is an
   effective (`cosmological') potential, and the coupling matrix of the
   scalar fields, $Z_{mn}(\varphi,\psi)$, is an arbitrary function of the
   dilaton fields and of the $(N-2)$ scalar massless fields $\psi_n$\footnote{
   In the models (\ref{aa.010}) derived by dimensional reductions of
   higher-dimensional (super)gravity theories the coupling matrix $Z_{mn}$
   is real, symmetric, and nondegenerate. We also suppose that all its
   eigenvalues are negative. The last condition may be violated if one
   introduces the so called `phantom' fields, but here we do not consider such
   exotic models. The usually present dilaton kinetic term
   $W (\nabla\varphi)^2$ can be transformed away from the Lagrangian and we
   omit it here (see \cite{ATF1}, \cite{ATF2}). }.
  This theory may be integrable in two main cases: 1) when $ V\equiv 0 $ and
  2) when all $Z_{mn}$ are constants independent of the fields, while $V$ is
  a special potential (see \cite{ATF2}, \cite{ATF3}).
   If $Z_{mn}$ has only negative eigenvalues we can transform
   $\psi$ fields in the last case so as
 \be
 \label{aa.030}
 Z_{mn}\,=\,-\delta_{mn}\,.
 \ee

Let us concentrate on this case with $V(\varphi,\psi)$ given by
\be
\label{aa.035}
V\,=\,\sum_{n=1}^N \,g_n {\rm
exp}\,{q_n^{(0)}}\,, \qquad q_n^{(0)}\equiv \,a_n\varphi\,+\,
\sum_{m=3}^N\,\psi_m \, a_{mn}\,.
\ee
The theory is integrable if
the parameters $a_n$, $a_{nm}$ satisfy the condition
 \be
 \label{aa.040}
\sum_{l=3}^N \, a_{lm} a_{ln}
\,-\,2\,(a_m\,+\,a_n\,)\,=\,\gamma_n^{-1} \, \delta_{mn}\,
 \ee
  and $Z_{mn}$ satisfy (\ref{aa.030}).
 When $m\not=n$ this is the
  pseudo orthogonality condition for the $N$-vectors $A_n\equiv
  \,\{a_{mn}\}$, $m=1,...,N$,
 $a_{1n} \equiv 1+a_n$, $a_{2n} \equiv 1-a_n\,$, i.e. (\ref{aa.040}) is
equivalent to
\be
 \label{aa.050}
 A_n \cdot A_m
  \, \equiv \, -\,a_{1m}\, a_{1m} \,+\, \sum_{l=2}^N\, a_{lm} a_{ln}
\,=\,0.
\ee
For $m=n$ eq.~(\ref{aa.040}) defines $\gamma_n$, the important parameters
of the theory,
 \be \label{aa.060}
  \gamma_n^{-1}\,=\, A_n \cdot A_n = - \,4 \,a_n \,
  + \,  \sum_{l=3}^N \, a_{ln}^2 \,  \,.
\ee
For a detailed description of the properties of the parameters $a_{mn}$,
$\gamma_n$ satisfying (\ref{aa.040}), see \cite{ATF1} -- \cite{ATF3}
   (there we showed that one and only one of the norms $\gamma_n^{-1}$
   is negative; usually we take $\gamma_1 <0$).

Note that the integrability does not depend on the values $g_n$ of the
   coupling parameters. Some of them may even be zero\footnote{
   If the theory (\ref{aa.010}) is obtained by dimensional reductions of
   a higher-dimensional supergravity theory on tori, the coupling constants
   $g_n$ are negative. An exception emerges in the case of the spherical symmetry
   reduction that produces one positive term
   in the two-dimensional potential (\ref{aa.035}).}.
 Also, without changing
integrability we may add to the theory (\ref{aa.010})   any number of free
massless scalar fields. In all these cases we can write explicitly the
general solution of the two-dimensional integrable theory defined by
equations (\ref{aa.010}) - (\ref{aa.040}).

In this paper we mainly concentrate on the dimensional reduction of the
(1+1)-dimensional integrable model to the one-dimensional integrable one.
Before turning to this main subject let us first outline relations of
these integrable models to realistic theories. To simplify presentation we
mainly work in the light-cone (conformally flat) gauge, where
 the metric is
\be
 \label{aa.070}
ds^2\,=-\,4\,f(u,v)\, du\, dv \,\equiv -4\varepsilon \,e^{F(u,v)}du\, dv
\,= \,e^F \, (dr^2\,-\,dt^2\,)\,, \,\,\, \varepsilon = \pm 1 \, ,
\ee
    with $r$ and $t$ defined as $r\equiv u - \varepsilon v$,
    $t\equiv u + \varepsilon v$. In the exponential representation of the
    metric in (\ref{aa.070}) the zeroes of $f(u,v)$ (horizons) may emerge when
    $F \rightarrow - \infty$. If we cross a horizon, $f$ changes sign
    given by $\varepsilon$ but the space-time representation remains
    unchanged due to the automatic changing of the relation between the
    space-time and the light-cone coordinates
    (resulting in $r \leftrightarrow$ t).

  The naive reduction to one dimension (space or time) is
obtained if we suppose that the metric, the dilaton, and the scalar matter
   fields depend a single variable $\tau=a(u)\,+\,b(v)$, which may be a space
or a time coordinate.
 The reduced Lagrangian then can be written in the form
 \be
 \label{aa.080}
 {\cal L}^{(1)}\,=\,-{1 \over l(\tau)}
    \, \biggl(\, \dot \varphi \, \dot F\, +\,\sum_{m,n=3}^N\, Z_{mn}
 {\dot \psi}_m \,{ \dot \psi}_n \,\biggr) \,+\, l(\tau)\,\varepsilon\,
 e^F\, V(\varphi,\psi)\,.
 \ee
If $Z_{mn}=-\delta_{mn}$, then eqs.~(\ref{aa.035}), (\ref{aa.040})  also
give the sufficient condition  for its integrability.

   However, the theory
   (\ref{aa.080}) may be integrable in other cases as well. For example,
   let us suppose that
   $Z_{mn} = - (1/\phi'(\varphi)) \, \delta_{mn}$,
   where $\phi'(\varphi)>0$. Then
\be
\label{aa.090}
{\cal L}^{(1)} \,=\, {1\over {l(\tau)\, \phi'(\phi)}}\,
 \biggl( -\dot  \phi \,\dot F \, + \, \sum_{n=3}^N \, {\dot \psi}_n^2  \,\biggr)
 \, + \, [l(\tau)\, \phi'(\varphi)] \,\varepsilon\,e^F \, [{V/ \phi'(\varphi)}]\,.
\ee
Introducing the new Lagrange multiplier
 $\bar l(\tau)\equiv l(\tau) \,\phi'(\phi)$ and defining
 \bdm
 V(\varphi,\psi)\,/\phi'(\varphi)\,\equiv\, {\bar V}(\phi,\psi)\,,
 \edm
 the new one-dimensional Lagrangian takes the form
 \be
 \label{aa.100}
 {\bar{\cal L}}^{(1)}\,=\, {1\over \bar l(\tau)}\,
 \biggl( - \dot \phi\, \dot F\, + \,\sum_{n = 3}^N\, \dot \psi_n^2 \,\biggr)
 \, + \, {\bar l(\tau)} \,\varepsilon \,e^F\,
 \bar V(\phi,\psi)\,.
 \ee
    This model is integrable if the new potential $\bar V(\phi,\psi)$
    can be written in the form (\ref{aa.035}) with the parameters
    $\bar a_n$, $a_{mn}$ ($m \geq 3$) satisfying the orthogonality
    conditions (\ref{aa.040}):
 \bean &&
 \label{aa.030'}
 \bar V(\phi,\psi)\,=\,\sum_{n=1}^N\, g_n\,{\rm exp}\,{\bar
 q_n^{(0)}}\,, \quad \bar q_n^{(0)}\,=\,\bar a_n\, \phi
 \,+\sum_{m=3}^N\, \psi_m \, a_{mn}\,,\\
&& \label {aa.040'}
   \sum_{l=3}^N \, a_{lm} \, a_{ln} \,-\,2\,
 (\bar a_m\,+\,\bar a_n\,)\, = \,\bar \gamma_n^{-1} \, \delta_{mn}\,.
 \eean

    Unlike the integrable one-dimensional model (\ref{aa.080}) obtained from
    the integrable two-dimensional model (\ref{aa.010}), both satisfying
    the integrability conditions (\ref{aa.030}) - (\ref{aa.040}), the new
    integrable model (\ref{aa.100}) is derived from the apparently
    nonintegrable two-dimensional theory. Indeed, to get the exponential in
    $\phi$ potential $\bar V$ we should start with the potential $V$
    non-exponential in $\varphi$  and with
    the $\varphi$-dependent $Z_{mn}$.
    To better understand this subtle relation between integrability in
    dimensions one and two,  consider a typical case
   $Z_{mn}=-\varphi\,\delta_{mn},\,\, \varphi= {\rm exp}\,\phi$.
   Then equations (\ref{aa.030'}), mean that (note that here
   $\bar a_n -1 = a_n$):
\bean
 \label{aa.110}
    V (\varphi , \psi)\, \equiv \, \phi'(\varphi) \, \bar V({\rm ln} \varphi,\psi)
    \, = \, \sum_{n=1}^N g_n \,\varphi^{\bar a_n -1} \,
    \exp \biggl[ {\sum_{m=3}^N \psi_m  a_{mn}} \biggr] \, \equiv
    \nonumber \\
   \equiv \,\sum_{n=1}^N g_n \,\varphi^{\bar a_n -1} \, e^{-\bar a_n \varphi} \,
     \exp \biggl[ {\bar a_n \varphi \, + \, \sum_{m=3}^N \psi_m a_{mn}} \biggr] \,,
\eean
   where $\bar a_n, a_{mn}$ satisfy the orthogonality relations (\ref{aa.040'}).

   Although the theory (\ref{aa.010}) with $Z_{mn}=-\varphi\,\delta_{mn}$
   and the potential (\ref{aa.110}) is certainly not integrable we may try
   to approximate it by an integrable, theory in some narrow enough interval of
   $\varphi$~\footnote{Such an approximation might be used in considering
   time-dependent phenomena near static horizons of black holes or,
   alternatively, when deriving an approximate description of
   nonperturbative short-time inhomogeneities in cosmology.}.
   First, we should approximate in this interval
   $Z_{mn} \approx -\varphi_0 \,\delta_{mn}$, where $\varphi_0 >0$.
   Second, we should either rescale the matter fields to have
   $Z_{mn} \approx - \delta_{mn}$ or simply divide all the terms in the
   Lagrangian (\ref{aa.010}) by $\varphi_0$ and then introduce
    $\bar \varphi \equiv \varphi / \varphi_0$ as the new
   dilaton field. The second
   idea looks more attractive because it preserves the
   pseudo orthogonality conditions for $\bar a_n, \, a_{mn}$ and rescales
   only the coupling constants $g_n$. After these simple manipulations we
   can forget about rescaling and simply fix the scale by setting
   $\varphi_0 = 1$. Then, to get the integrable two-dimensional theory
   we suppose that
   $\bar g_n (\varphi) \equiv g_n \varphi^{\bar a_n-1} \, e^{-\bar a_n \varphi}$
   can be approximated by constants in some interval of
   $\varphi$ around $\varphi_0$. The simplest approximation is to replace
   in this expression $\varphi$ by $\varphi_0$. A more refined
   approximation may be possible if several parameters $\bar a_n$ are
   equal, say, $\bar a_n = \bar a$ for $n \leq n_0$. If
   $g_n = 0$ for $n_0 < n \leq N$ and thus the potential has $n_0$ terms,
   we can use a better approximation for the potential.
   The derivative of $g_n \varphi^{\bar a_n-1} \, e^{-\bar a_n \varphi}$
   vanishes if $\varphi = \varphi_1 \equiv (\bar a_n-1) / \bar a_n$.
    This means that, if $|\varphi-\varphi_1|$ is not too large, the
   approximation of $\bar g_n (\varphi)$ by $\bar g_n (\varphi_1)$
   is reasonable for $n \leq n_0$ and therefore
\be
\label{110a}
 V(\varphi,\psi)\,=\, \sum_{n=1}^{n_0}\, \,g_n (\varphi)
   \exp \biggl[{\bar a \varphi +\sum_{m=3}^N \psi_m a_{mn}} \biggr]
 \approx \, \sum_{n=1}^{n_0}\, \bar g_n (\varphi_1) \,
 \exp \biggl[{\bar a \varphi + \sum_{m=3}^N \psi_m a_{mn}} \biggr] .
\ee

     The simplest examples of one-dimensional
     integrable $N$-Liouville models that can be
   obtained by dimensional reductions of higher-dimensional theories are:
   $\bar a_1 = \bar a, \bar a_n = -\bar a, n>1$ and
   $\bar a_1 = \bar a_2 = \bar a, \bar a_n =-\bar a, n>2$.
   They can be derived from the models considered in detail in \cite{Kiem}
   (see also \cite{Lid} and Appendix for more general models).
   To better understand the above discussion it is advisable to have a
   look at the model of ref.~\cite{Kiem}, which we present in our notation
   and without reproducing the higher-dimensional considerations.
   The two-dimensional Lagrangian can be written in the form
   (\ref{aa.010}) with $N=4$, $Z_{mn} = -\varphi \, \delta_{mn}$ and
   with the potential (\ref{aa.110}), in which $\bar a_n, a_{mn}$ satisfy
   (\ref{aa.040'}). The nonzero parameters of
   the model are:
\be
   \bar a_1 = \bar a_2 = -\bar a_3 = -\bar a_4 = 1; \,\,\,
   a_{31} = 2\sqrt 3, \,\,
   a_{32} = 2 / \sqrt 3, \,\, a_{43} = -a_{44} = 2 \, .
\label{110b}
\ee
   The origin of the items in the potential is the following: the first
   term is the curvature of the three-dimensional sphere, $g_1 = 6$
   (in the flat limit this term vanishes), the second and the fourth terms
   originate from the reductions of the three-form in the ten-dimensional
   Lagrangian, the third is generated by the Abelian gauge field generated
   by the KK-reduction of the six-dimensional space-time to the five-dimensional
   one; $\psi_4$ is the KK scalar field and
   $\psi_3$ is the scale factor of the five-dimensional spherical
   reduction. The coupling constants $g_2, g_3, g_4$ are negative
   (note that $g_n / \gamma_n > 0, \,\, n=1,...,4$).
   We can now write the one-dimensional Lagrangian (\ref{aa.080}) and
   transform it into the $N$-Liouville form (\ref{aa.100}). The described
   special model and its generalizations can be used as a `laboratory'
   for confronting exact and approximate solutions.

   We therefore see that in the `realistic' case, $Z_{mn}=-\varphi\,
\delta_{mn}$, the one-dimensional theory may be reducible to the
$N$-Liouville model while the two-dimensional theory, from which it was
   obtained by the naive dimensional reduction, is not integrable. Although we
know that the one-dimensional solutions must also solve the
two-dimensional equations, the procedure of finding the corresponding
two-dimensional functions requires some care because we used
transformations of Lagrange multipliers (in this sense the reduction is
not `naive' at all!). Moreover, we do not know the complete set of
solutions of the two-dimensional theory and
  thus cannot directly dimensional reduce of the solutions
  (as it is possible in the integrable case).

Having in mind that the `effectively one-dimensional solutions' formally
depending on two variables are rather difficult to find in the
  nonintegrable theory, we first carefully study the integrable model,
\be
 \label{aa.140}
{\cal L}^{(2)}\,=\,\sqrt{-g}\,\biggl\{ \, \varphi\,R \,+\,
 V(\varphi,\psi)\, -\sum_{n=3}^N\,(\nabla \psi_n)^2 \,\biggr\} \, ,
\ee
with $V(\varphi, \psi)$ given by eqs.~(\ref{aa.030}) - (\ref{aa.040}), and
study the dimensional reduction of its solutions to static states,
cosmologies and waves. In view of the connections between the
one-dimensional and the two-dimensional theories in `realistic' cases, one
may hope that these considerations will help to find analogous reductions
in realistic theories by using a generalized dimensional
  reduction (in Section~5, we give a simple example hinting that this should be
  possible). To substantiate this observation, we also show how the
wave-like solutions of the $N$-Liouville theory (\ref{aa.010}) -
   (\ref{aa.040}) can be obtained with the aid of a generalized separation
   of variables in the equations of motion, without constructing the general
two-dimensional solutions.

\section{General solution of the $N$-Liouville model}

We have shown in \cite{VDA1} -- \cite{ATF3} that the equations of motion
for the $N$-Liouville model defined by (\ref{aa.010}) -- (\ref{aa.040})
greatly simplify if we write them in the light-cone (LC) coordinates
(\ref{aa.070}) using the following new functions
\be
\label{001}
   q_n(u,v) \equiv F(u,v) \,+ \, q_n^{(0)}(u,v) \,
   = \sum_{m=1}^N \psi_m a_{mn} \, ,
\ee
  where $\psi_1 \equiv \half (F+\varphi)$, $\psi_2 \equiv \half (F- \varphi)$
  and $a_{mn}$ are the coordinates of the $N$-vectors $A_n$ defined above.
  Using (\ref{aa.050}), (\ref{aa.060}) it is easy to invert (\ref{001})
  and express $\psi_n$ in terms of $q_n$ (see \cite{ATF4}, \cite{ATF2})
\be
\label{001a}
   \psi_n \, = \, \sum_{m=1}^N \epsilon_n a_{nm} \gamma_m \, q_m \,\,\,\, ,
   \epsilon_1 = -1, \,\, \epsilon_{n>1} = 1 \, .
\ee

  Varying the Lagrangian (\ref{aa.140}) in $\psi_n$, $\varphi$, $g_{ij}$
  and then transforming the obtained equations to the LC coordinates,
  we can find $N$ equations of motion for $\psi_{n \geq 1}$ and two
  constraints. Expressing $\psi_n$ in terms of $q_n$ we thus obtain
  $N$ Liouville equations\footnote{
  In physically motivated models, e.g., in those obtained from
  higher-dimensional theories by dimensional reductions, the signs of $g_n$ and of
  $\gamma_n$ are often correlated so that the signs of
  nonvanishing coupling constants $\tilde g_n$ are the same
  for all $n$.},
\be
\label{002}
\p_u \p_v q_n = \tilde g_n \exp{q_n} , \,\,\,\, \tilde g_n \equiv
\varepsilon g_n / \gamma_n \, ,
\ee
and two constraints for $n$ functions $q_n$,
\be
\label{003}
\sum_{n=1}^N \gamma_n \biggr[(\p_i q_n)^2 - 2 \p_i^2 q_n \biggl] =0 .
\ee

The solution of the system of equations
  and constraints is easier to find if we define the new functions
  (motivated by the conformal properties of the Liouville equation)
\be
\label{004}
X_n(u,v) \equiv \exp \bigr[- q_n (u,v)/2 \bigl] \, .
\ee
Then the Liouville equations (\ref{002}) are equivalent to
\be
\label{01}
X_n \,\p_u \p_v X_n  - \p_u X_n \,\p_v X_n  = -\half \tilde g_n
\ee
and their solution
must
satisfy the constraints
\be
\label{005}
\sum_{n=1}^N \gamma_n\,X_n^{-1}\,\p_i^2 X_n \,=\,0 , \,\,\, i = u,v .
\ee
The solution of
 these equation is described in detail in \cite{VDA1} --
\cite{ATF3} and we will not repeat it. Here we only write a more general
form of the solution that we need for a correct dimensional
reduction\footnote{The form used in the previous papers is essentially
correct but was written in a somewhat misleading form that can provoke its
incorrect use; this will become more evident below.}.

Now let us proceed to writing the general solutions of the $N$-Liouville
theory.   Differentiating eq.~(\ref{01}) with respect to $u$ and $v$ we
find that
\be
\label{02}
\dd_u \bigl( X_n^{-1} \, \dd_v^2 X_n \bigr) \,=\,0 \, ,
\qquad
\dd_v \bigl( X_n^{-1} \, \dd_u^2 X_n  \bigr)\,=\,0 \,.
\ee
It follows that if $X_n$ satisfies eq.~(\ref{02})  then there exist some
`potentials' ${\cal{U}}_n(u)$, ${\cal{V}}_n(v)$ such that
\be
\label{03}
\dd_u^2 X_n \,-\,{\cal U}_n(u) \, X_n\,=\,0, \qquad
\dd_v^2 X_n \,-\,{\cal V}_n(v) \, X_n\,=\,0 \,.
\ee
This motivates the introduction of two ordinary differential equations
\be
\label{03a}
a''_n(u) \, - \,{\cal U}_n(u) \, a_n(u) \, = \,0,
\qquad
b''_n(v)  \, - \, {\cal V}_n(v) \, b_n(v) \, = \, 0 \, .
\ee
   It follows that the solutions of the
Liouville equations, $X_n$, can be expressed in terms of the solutions of
these equations: $a_n(u)$ and $b_n(v)$ can be taken arbitrary and the
linearly independent solutions are then found without knowledge of the
   potential\footnote{This form for the general solution of the Liouville
   equation was mentioned in \cite{DPP}. It group theoretical meaning and
   applications are discussed in  \cite{Gervais}, \cite{Leznov}, where
   further references can be found. }.
   However, this does not help to solve the constrains
(\ref{005}) that will be shown below to have the form
\be
\label{12}
\sum_{n=1}^N \, \gamma_n\, {a^{''}_n(u) \over a_n(u) } \,= \,0 ,
   \,\,\,\,\,\,\,\, \sum_{n=1}^N \, \gamma_n\, {b^{''}_n(v) \over b_n(v)}
   \, = \,0 \, ,
\ee
where $a_n(u)$ and $b_n(v)$ are arbitrary solutions of eqs.~(\ref{12}).
Although $2(N-1)$ of the potentials ${\cal U}_n(u)$, ${\cal V}_n(v)$ and
of the corresponding $a_n(u)$, $b_n(v)$ are thus arbitrary, the two
remaining functions $a$, $b$ satisfy the second order equations with
arbitrary potentials that cannot be solved when $\gamma_n$ are arbitrary
numbers. In \cite{VDA1} -- \cite{ATF3} we showed that using the fact that
in the $N$-Liouville theory the sum of $\gamma_n$ is zero the solution can
be constructed. Below we slightly generalize this result\footnote{For some
special potentials, the equations (\ref{03a}) can be solved even for
arbitrary $\gamma_n$. Below we study in detail the case of constant
potentials that can be considered as a sort of a generalized dimensional
reduction of the $N$-Liouville theory with or without constraints.}.

 Introducing the linearly independent
solutions of eqs.~(\ref{03a}), $a^{(1)}_n(u)$, $a^{(2)}_n(u)$ and
$b^{(1)}_n(v)$, $b^{(2)}_n(v)$, normalized by requiring that their
wronskians be
\be
\label{03b}
W\biggl[a^{(1)}_n(u), \,a^{(2)}_n(u) \,\biggr]\,=\,1, \qquad
W\biggl[b^{(1)}_n(v), \,b^{(2)}_n(v) \,\biggr]\,=\,1\,,
\ee
we may show that the $X_n$ satisfying eq.~(\ref{02}) should have the form
\be \label{04}
X_n(u,v)\,=\, a^{(i)}_n(u) \, C^{(n)}_{ij} b^{(j)}_n(v)
\ee
where $C_{ij}^{(n)} $ is a numerical matrix (and we sum over $i,j = 1,2$).
It  is not difficult to check that $X_n$ satisfies eqs.~(\ref{01}) if and
only if
\be \label{05}
   \det \,C_{ij}^{(n)} \,=\, - \half \tilde g_n  \,.
\ee

In \cite{VDA1} -- \cite{ATF2} we used a truncated form of the solution
obtained taking $C_{12}^{(n)}=C_{21}^{(n)}=0$, $C_{11}^{(n)} \,= \,1$ and
$C_{22}^{(n)} \,= \,-\tilde g_n/2$. The general solution can be obtained
from the truncated one by applying the linear transformations of the basic
chiral fields $a_n^{(i)}(u)$, $b_n^{(j)}(v)$  that preserve the
wronskians:
\be
\label{06}
a_n^{(i)}(u)\,\ra A_{ij}^{(n)} \, a_n^{(j)}(u), \quad
  b_n^{(i)}(v)\,\ra B_{ij}^{(n)} \, b_n^{(j)}(v),
\ee
where $\det A_{ij}^{(n)} \,=\,1\,=\,\det B_{ij}^{(n)} $. Although the
truncated representation of the solution can be, with due care, used in
general considerations, the general solution (\ref{04}) is more adequate
in order to analyze the physical properties of the solutions and is
     especially important for further
     dimensional reductions as we shall see below.

As in the previous papers, we may choose $a_n^{(1)}(u) \equiv a_n(u)$ and
$b_n^{(1)}(v)\equiv b_n(v)$, where $a_n(u)$ and $b_n(v)$ are arbitrary
functions, and take
\be
  a_n^{(2)}(u) \,\equiv \,\bar a_n(u)\,\equiv \, a_n(u)\,\int\,
 {du \over a_n^2(u) } \,, \,\,\,\,\,\,\,\,
  b_n^{(2)}(v) \,\,\equiv \,\,\bar b_n(v)\,\equiv \,\, b_n(v)\,\int\,
 {dv \over b_n^2(v) }\,.
\label{07}
\ee
This allows us to apply the considerations of the previous papers to the
general solutions $X_n$. The crucial fact that makes it possible to solve
the constraints in \cite{VDA1} -- \cite{ATF2} is that they are equivalent
to the constraints (\ref{12}) and that the pseudo orthogonality conditions
imply the identity
\be
\label{12a}
\sum_{n=1}^N \, \gamma_n \, = \, 0 \, .
\ee
Indeed, we have
\bdm
{\dd^2_u X_n \over X_n} \,=\, {\bigl[\dd^2_u \,\,a_n^{(i)}(u)\,\bigr]\,
C^{(n)}_{ij} \, b^{(i)}_n(v)\, \over a_n^{(i)}(u) \, C_n^{(ij)} \,
b_n^{(j)}(v) } = {\cal U}_n(u) = {a_n^{''}(u) \over a_n(u) } \,=\,{{\bar
a}_n^{''}(u) \over {\bar a}_n(u) } ,
\edm
and a similar expression holds for the functions $b_n(v)$,
${\bar b}_n(v)$.
It follows that the constraints (\ref{005}) have the form (\ref{12}) that has
allowed us to solve the constraints in \cite{VDA1} -- \cite{ATF2}. Now
applying the construction of these papers we can write the solution of the
equations and constraints with the same expression for the basic functions
$a_n$ and $b_n$:
\be
\label{13}
a_n(u)\,=\, |\sum\gamma_m\mu_m(u)|^{-1/2} \, \exp\,\int\,du\, \mu_n(u),
\ee
\be
\label{13a}
b_n(u)\,=\, |\sum\gamma_m\nu_m(v)|^{-1/2} \, \exp\,\int\,du\, \nu_n(v),
\ee
where the moduli functions $\mu_n(u),$ $\nu_n(v)$ satisfy the
constraints\footnote{It is easy to write the potentials ${\cal U}_n(u)$,
${\cal V}_n(v)$ in terms of the moduli but these complex formulas do not
allow to find  inverse expressions except the case of constant moduli
treated in the next Section.}
\be
\label{10}
   \sum_{n=1}^N \gamma_n \, \mu_n^2(u)\,=\,0 \,, \,\,\,\,\,\,\,\,\,
   \sum_{n=1}^N \gamma_n  \,=\,0 \,.
\nu_n^2(v)\,.
\ee
Inserting (\ref{13}), (\ref{13a}) into (\ref{04}) we find $q_n$ and then
using the orthogonality relations the original fields $\psi_n$, $\phi$,
$F$ can be written in terms of $a_n$, $b_n$ (see \cite{VDA1} --
\cite{ATF2}; these expressions are reproduced in the next Section).

As it has been shown in refs. \cite{ATF1}, \cite{ATF2}, the space of
moduli $(\mu_1,\, ...,\mu_n)$, $(\nu_1,\, ...,\nu_n)$ may be reduced
further by the two coordinate gauge fixing conditions. In \cite{ATF1},
\cite{ATF2} it was suggested to introduce the new coordinates $(U,\,V)$ by
writing
\be
\label{14}
\sum \gamma_n \mu_n(u)\,=\,U'(u),\qquad \sum \gamma_n \nu_n(v)\,=
\,V'(v)\,,
\ee
  but in some problems a different choice of coordinate
conditions may become more convenient.

In the mentioned papers it was shown that the pseudo orthogonality
conditions imply that all $\gamma_n$ except one should be positive, so we
always choose $\gamma_1<0$. It follows then that $\mu_1(u)$ and $\nu_1(v)$
cannot have zeroes, due to the constraints\footnote{This is not true if
all $\mu_n$ or $\nu_n$ vanish. Then the constraints are trivially
satisfied but this highly degenerate case is not interesting to discuss.
More interesting is the case of imaginary moduli, to which our approach
can be fully applied. We do not consider the corresponding solutions
because they may have singularities for finite values of $u$ and $v$. In
this paper we concentrate on solutions that can be singular only at
infinity in the ($u,v$)-space. }. Accordingly,
\be
\label{15}
\int_{u_0}^u\,\mu_1(u)\,du \, \qquad {\rm and} \qquad \,
\int_{v_0}^v\,\nu_1(v)\,dv
\ee
are monotonic functions and so they can be chosen as new coordinates, i.e.
we may write
\be
\label{16}
 \int_{u_0}^u\, \mu_1(u) \,du\,\equiv\,U\, \quad  {\rm and}  \quad
 \int_{v_0}^v \, \nu_1(v)\,dv\,\equiv V\,.
\ee
This is obviously equivalent to the choice
\be
\label{17}
\mu_1(u)\,\equiv \, \nu_1(v) \, \equiv \,1 \,
\ee
while using the original coordinates $(u,\,v)$.

A simple way  to incorporate all the above properties of the moduli
space is to introduce the unit vectors (see \cite{ATF1}, \cite{ATF2})
\be
\label{18}
\hat \xi_k(u) \, \equiv {\hat \gamma_k} \,\, \mu_k(u) , \qquad \hat
 \eta_k(v) \, \equiv\, {\hat \gamma_k} \,\, \nu_k(v) , \quad k=2,3,...,N \,,
\ee
where $\hat \gamma_k \equiv \, \sqrt {\gamma_k/|\gamma_1|}$. From the
known properties  of the parameters $\gamma_k$ and the constraints on
$\mu_n$ and $\nu_n$ (including $\mu_1= \nu_1=1$) we see that
\be
\label{19}
{\hat \xi}^2\,\equiv \,\sum_{k=2}^N \, {\hat \xi_k}^2\,=\,1,\quad
 {\hat \eta}^2 \,\equiv\,\sum_{k=2}^N \, {\hat \eta_k}^2\,=1, \quad
\hat \gamma^2\,\equiv\,\sum_{k=2}^N\,{\hat \gamma_k}^2\,=\,1\,.
\ee

The independent moduli $\hat \xi(u),\, \hat \eta(v)$ belong to a sphere
$S^{(N-2)}$ and the moduli space consists of pairs of continuous
trajectories $(\hat \xi(u)$, $\hat \eta(v))$ on
   $S^{(N-2)}$. Pairs of points ($\hat \xi^{(0)}$, $\hat \eta^{(0)}$)
also define an interesting class of solutions that will be studied below
  in some detail\footnote{In our previous papers we mainly studied the
  solutions for which $\hat \xi^{(0)} = \hat \eta^{(0)}$. The static
  solution with two horizons are described by the solutions
  defined by $\hat \xi^{(0)} = \hat \eta^{(0)} = \hat \gamma$
  (see \cite{VDA1}).
  These are very special examples of the solutions discussed in the next
  Section. }.
  The moduli $\hat \xi(u)$ and $\hat \eta(v)$ should be
defined for all the real values of the coordinates $(u,\,v)$. These
trajectories in $S^{(N-2)}$ may form closed curves or be infinite in both
directions or even have one or two finite end points. If a trajectory, say
$\hat \xi(u)$, has an end point $\hat \xi^{(0)}$, this means that $\hat
\xi(u) \ra \hat\xi^{(0)} $ for $u \ra +\infty$ (similarly it can have
$\hat \xi^{(0)}$  as initial point if $\xi(u)\ra \hat \xi^{(0)}$ for $u\ra
-\infty$).

Using these simple properties one may try to construct a classification of
trajectories by their topological and asymptotic properties. This is
fairly obvious for $S^{(1)}$ and simple enough for $S^{(2)}$. In higher
dimensions (starting with $S^{(3)}$) it may be rather difficult to find a
reasonable classification of all possible trajectories. However, many
physically interesting solutions are those defined by the points
 $(\hat \xi^{(0)},\,\hat\eta^{(0)})$.
 They, in fact, define some nontrivial dimensional
reductions of the 1+1 dimensional solutions and may describe static black
holes (more generally, static states of gravitating matter), cosmologies
and various types of nonlinear waves coupled to gravity. Among these waves
   there may exist waves that are finite everywhere in space and time,
   including infinity. All these solutions
   will be one of the main subjects of the following discussion.

Let us first make some comments about the dimensional reduction, extending
   the discussion on this subject given in two previous papers
   \cite{ATF1}, \cite{ATF2}. In the `naive' reduction we simply reduce
   the dimension of
space-time on which the dynamical functions (say, $q_n(u,\,v)$) are
defined, by constructing a subclass of solutions dependent on one variable
(e.g. $q_n(t)\,\equiv \, q_n(u-v)\,)$. This naive reduction for the 1+1
dimensional dilaton gravity theory functions is discussed in detail in
\cite{ATF1}. There we also argued that the naive reduction to variables
dependent only on time  does not give all possible cosmologies. It also
looks as if the reduction from higher dimensions $D$ to 1+0 is not
equivalent to the reduction from dimension 1+1 to 1+0. In appendix 6.2 of
paper \cite{ATF1} we show how to restore all possible homogeneous and
isotropic cosmologies by considering a more general reduction of the 1+1
dimensional theory.

More general approaches to reductions use group-theoretical considerations
(see e.g. \cite{Fre} - \cite{Iv} and references therein) but these ones
are not always easy to apply directly to the reduction of the equations of
motion. The generalized separation of variables introduced in \cite{ATF3}
may give an even more general approach to dimensional reduction. In
particular, when applied to theories of dimension 1+1, this approach
allows one to reproduce all known black hole and cosmological solutions
\cite{ATF3}. In addition, it works well  when the dilaton gravity couples
to any number of matter fields and gives (presumably) new solutions that,
while essentially one-dimensional, depend on both variables, $r$ and $t$.

Here we study  the dimensional reduction in the integrable $N$-Liouville
theory that gives a class of reduced solutions dependent on combinations
of both variables. An example of this type of solution is given at the end
of Section 3 of ref. \cite{ATF2}; it was obtained by dimensional reduction
of the moduli space. The idea of this dimensional reduction (let us
temporary call it a `dynamical dimensional reduction') is very simple to
formulate in terms of the moduli space described above. While the
solutions of the 1+1 dimensional theory are given by the
  pairs of functions, ($\hat \xi(u)$, $\hat \eta(v)$),
  the reduced solutions are given by the points
($\hat \xi^{(0)}$, $\hat \eta^{(0)}$). They describe waves of scalar
matter; special cases are given by static states (black holes in
 particular) and cosmologies. Therefore all these objects are unified in
 one more general class.

We shall study them in the next Section and will try to find out those
ones that have matter fields remaining finite at both $r\ra \infty$ and
$t\ra \infty$. Then we will demonstrate that instead of this dynamical
dimensional reduction we can construct the same states by using the
generalized separation of the space and time variables. This is a fairly
simple exercise that does not require knowledge of the (1+1)-dimensional
solutions. Thus it can in principle be applied to nonintegrable problems.
We discuss this at the end of the paper using a very simple example.

\section{States with constant moduli}

   Let us consider solutions with constant moduli $\mu_n, \, \nu_n$.
   Then  ${\cal U}_n(u) = \mu_n$, ${\cal V}_n(u) = \nu_n$ and it is easy to write
explicitly the basic chiral fields. If no moduli vanish\footnote{If some
   $\mu_n$ are allowed to vanish we may use different basic functions, e.g.,
    $a_n = \cosh {\mu_n}$, $\bar a_n = \mu^{-1}_n \sinh {\mu_n}$, but we shall not discuss
   these subtleties here.}
 we can use the following set according to conditions (\ref{03a}):
\be
\label{c1}
a_n(u)\,={1\over \sqrt{2 \,\mu_n}}\, e^{-\mu_nu}, \quad
b_n(u)\,={1\over \sqrt{2 \,\nu_n}}\, e^{-\nu_nu},
\ee
\be
\label {cc1}
\bar a_n(u)\,=  {1\over \sqrt{2 \,\mu_n}}\, e^{\mu_nu}, \quad \bar
b_n(u)\,={1\over \sqrt{2 \,\nu_n}}\, e^{\nu_nu}.
\ee
Then the general constant - moduli solution of eqs.~(\ref{01}) may be
written as
\bean
\label{c2}
X_n(u,v)\,=&& \,{1\over 2\sqrt{\mu_n \nu_n}}\, \biggl[\,C_{11}^{(n)}
\, e^{-\mu_nu-\nu_nv} \,+\, C_{22}^{(n)}\, e^{\mu_nu+\nu_nv}\,+
\nonumber \\
&&+\,C_{12}^{(n)} e^{-\mu_nu+\nu_nv}+C_{21}^{(n)} e^{\mu_nu-\nu_nv}
\,\biggr] \,,
\eean
where
\be
\label{c3}
   C_{11}^{(n)}\,C_{22}^{(n)}\, -\, C_{12}^{n)}\, C_{21}^{(n)}\, =\,
   -\half \tilde g_n \,.
\ee

In general, when all the elements $C_{ij}^{(n)}\not= 0$, we may, without
loss of generality, regard all the moduli to be non-negative, $\mu_n \geq
0$, $\nu_n \geq 0$. If they are not all vanishing, then $\mu_1\not= 0$,
$\nu_1\not=0$ due to the constraints (recall that $\gamma_1 < 0$ and
$\gamma_k > 0$)
\be
\label{c4}
|\gamma_1| \mu_1^2 \,=\,\sum_{k=2}^N \gamma_k \mu_k^2, \quad
 |\gamma_1| \nu_1^2 \,=\,\sum_{k=2}^N \gamma_k \nu_k^2\,.
\ee
To make our presentation more concise, in what follows we suppose that
$\mu_n>0$, $\nu_n>0$. Recall also that we can choose the gauge in which
$\mu_1 = \nu_1 =1$. In this gauge we have
   $\mu_k \leq \sqrt{|\gamma_1|/\gamma_k}$, $k>1$.

The matrix elements $C_{ij}^{(n)}$ may in general have different signs and
thus $X_n(u,v)$ may have zeroes at some finite points
 $(u_0, v_0)$~\footnote{In what follows we usually suppose that
 $-\infty  < u < +\infty$, $-\infty < v < +\infty )$.},
 $X_n(u_0,v_0)=0$, and correspondingly
 $q_n(u_0,v_0) \,=\,-2 \ln \,|X_n(u_0,v_0)| \,=\, \infty$.
   In general, at these points the matter fields will be infinite too.
   In order to
   avoid these singularities we suppose that all
$C_{ij}^{(n)}>0$~\footnote{The existence of singularities is a typical
feature of the Liouville equation that was carefully studied by different
   researchers in the past, see, e.g.,
   \cite{DPP}, \cite{Polivanov}, \cite{Jackiw}. As
explained in Introduction, our purpose is to study minimally singular
solutions and thus we allow for singularities at infinity only.}. Of
course, special solutions with $C_{12}^{(n)}= C_{21}^{(n)} =0$ or with
$C_{11}^{(n)}=C_{22}^{(n)}=0$ for all $n$ are also of interest because, as
will be clear in a moment, they may describe static or cosmological
states. Note that our remarks on singularities are fully applicable to the
general solution (\ref{04}). To avoid singularities at finite points we in
general should suppose that the basic chiral functions are positive and
  $C_{ij}^{(n)} \geq 0$ (some of them should not vanish).

With positive $C_{ij}^{(n)}$, it is natural to rewrite the solution as
\be
\label{c5}
X_n={1\over \sqrt{\mu_n \nu_n}} \biggl\{ C_n^+\, \cosh \biggl(\mu_nu+\nu_nv +
\delta^+_n \biggr) \,+\,C_n^-\, \cosh \biggl(\mu_n u-\nu_n v+\delta_n^-
\biggr) \,\biggr\}
\ee
where we define $C_n^{\pm}$, $\delta_n^{\pm}$ by the equations
\be \label{c6a}
C_n^+\,=C_{11}^{(n)} \, e^{\delta_n^+} \,=\,C_{22}^{(n)}\, e^{-\delta_n^+},\,\
C_n^-\,=C_{12}^{(n)} \, e^{\delta_n^-} \,=\,C_{21}^{(n)}\, e^{-\delta_n^-},
\ee
\be
\label{c6b}
\delta_n^+={1\over 2} \, \ln { C_{11}^{(n)} \over C_{22}^{(n)}}, \quad
\delta_n^-={1\over 2} \, \ln { C_{12}^{(n)} \over C_{21}^{(n)}}
\ee
 and, according to (\ref{c3}), we have
\be
\label{c6}
 \biggl(C_n^+ \biggr)^2\, -\,\biggl(C_n^- \biggr)^2\,=\,-
  \half \tilde g_n \,.
\ee
   Note the important property of this simple formula: $C_n^-  =  0$
   is only possible if $\tilde g_n < 0$ while  we can
   take $C_n^+  =  0$  for $\tilde g_n > 0$ only. Recalling that
   $\tilde g_n \equiv \varepsilon g_n / \gamma_n$, where $\varepsilon$
   is the sign of the metric $f$, we infer the following fact:
   as $\gamma_1 < 0$ and $\gamma_k > 0$ for $k>1$, the common sign of all
     $\tilde g_n$ is possible only if $g_1 / g_k <0$ (or $g_1 =0$). This
     is important for constructing the standard one-dimensional solutions.

   The static or cosmological solutions can be obtained
from (\ref{c5}) by taking $\mu_n = \nu_n$ and $C_n^-=0$ or $C_n^+ = 0$ for
all $n$. Then the static states depend on the space coordinate defined as
 $r \equiv (u -\varepsilon v)$ while the cosmological states depend on the time
coordinate
 $t \equiv (u +\varepsilon v)$~\footnote{Of course, the real identification of the space
 and time coordinates in dimension (1+1) is only possible if we relate the
 two-dimensional metric to a higher-dimensional one, see a discussion in
   \cite{ATF1}. In what follows we usually take $r = u+v$ and $t=u-v$ having in mind
   that all formulas are essentially symmetric under the
   substitution $r \leftrightarrow t$}.
 Note that in order to obtain
 all possible static and cosmological
solutions from (\ref{c2}) we should allow not just positive matrix
elements $C_{ij}^{(n)}$. Note also that in general one may take imaginary
$\mu_n$ and $\nu_n$ keeping $X_n(u,v)$ real but in this paper we will not
consider oscillating waves that necessarily have singularities at finite
points of the $(u,v)$ space.

For a better understanding of the physical meaning of these solutions we
rewrite them as functions of $r$ and $t$:
\be
\label{c5a}
X_n = {1\over \sqrt{\mu_n \nu_n}}
 \biggl\{ C_n^+\, \cosh \biggl(\lambda_n \,r +
 \bar \lambda_n \, t \, + \delta_n^+ \bigg) \, + \,
 C_n^- \, \cosh \biggl(\lambda_n \, t +
 \bar \lambda_n \, r + \delta_n^- \bigg) \,\biggr\} \, ,
\ee
where we use the notation $\lambda_n \,=\,(\mu_n + \nu_n)/ 2$ and
 $\bar \lambda_n\,=\, (\mu_n - \nu_n)/ 2$.
 We see that $X_n$ are sums of two waves having the phase velocities
 $V_n^- \equiv \lambda_n / \bar \lambda_n$ and $V_n^+ \equiv
 \bar \lambda_n / \lambda_n$.

Now we turn to the study of the asymptotic behaviour of the fields
$\psi_m$, $\vf$, $F$ for the class of solutions given by  (\ref{c5}).
First, we write and briefly discuss the general formulas.  As it follows
from equations (46), (47) of ref.~\cite{ATF2}, the expressions for the
original fields in terms of $X_n$ are the following:
\bean
 && \psi_m\,=\,-2\,\sum_{n=1}^N\, a_{mn}\gamma_n \,\ln \,|X_n(u,v)|,
 \quad m\geq 3,
\label{c8c} \\
 && \vf=4\sum_{n=1}^N\, \gamma_n \,\ln \,  |X_n(u,v)|,
\label{c8b}\\
 && F=4\sum_{n=1}^N\, a_n\gamma_n \,\ln\, |X_n(u,v)|,
\label{c8a}
\eean
and the curvature of the 1+1 dimensional space-time is
\be
\label{c9}
 R\, = \,e^{-F}\, \dd_u\dd_v F\,= \, -\Pi_{n=1}^N\,
 |X_n(u,v)|^{-4a_n\gamma_n} \, \cdot \,
 \sum_{m=1}^N2 a_m \gamma_m \, \tilde g_m\, X_m^{-2}
\ee
where we have used eq.~(\ref{01}).

As we emphasized above, we suppose that the solutions have no
singularities for finite points $(u,v)$. In general, they have
singularities when the variables $(u,v)$ (or $(r,t)$) are infinite.
However, for some choices of moduli the infinities may cancel out. For
example, if all the functions $X_n$ are infinite and the quantities
 $\ln |X_n(u,v)|$ have the same asymptotic behaviour, i.e.
\bdm
\ln |X_n(u,v)| = \ln |X(u,v)| + f_n(u,v) ,
\edm
where $f_n(u,v)$ are asymptotically finite, then $\psi_m$ and $\vf$ may be
finite and $F \ra -\infty$ (this means that $|f|=e^F\ra 0$), as it can be
seen from the relations (see \cite{ATF1}, \cite{ATF2})
\be
\label{c10}
\sum_{n=1}^N\, \gamma_n\,=\,0; \quad \sum_{n=1}^N\, a_{mn}\gamma_n=0,\,\,
m \geq 3; \quad 4\sum_{n=1}^N\,a_n\gamma_n=-2 \, .
\ee
It follows  that also $R$ in this case is finite. In fact, as it has been
shown in \cite{ATF1}, \cite{ATF2}, this behaviour may be realized for
dimensionally reduced, static (depending only on $r$) solutions, which for
$r\ra \pm \infty$ have horizons as $|f| \rightarrow 0$.

Now let us turn to the general dimensionally reduced solution (\ref{c5})
with constant $\mu_n$ and $\nu_n$. We look for minimally singular
solutions that, being finite at space and/or time infinity, can be
regarded as localized in some weak sense. To study the singularities at
infinity of the solutions with constant moduli, we can use the general
formulas (\ref{c8c}) - (\ref{c8a}) to find their asymptotic behaviour for
$r\ra \infty$ and/or $t\ra \infty$. It is not difficult to find
   the following asymptotics
\be
\label{c11}
   \ln \, |X_n(u,v)|_{r\ra \pm \infty} = \pm (\lambda_n \, r +
   \bar \lambda_n \, t \, + \delta_n^+ ) +
   \ln {C_n^+\over 2\sqrt{\mu_n \nu_n}} \, + \, ... \,,
\ee
\be
\label{c12}
   \ln \, |X_n(u,v)|_{t \ra \pm \infty} = \pm ( \lambda_n \, t +
   \bar \lambda_n \, r + \delta_n^- ) +
   \ln {C_n^-\over 2 \sqrt{\mu_n \nu_n}} \, + \, ... \,.
\ee
The omitted terms are exponentially small in the two cases:\\
1) when $r\ra \pm \infty$ and $t$ is finite or also $t\ra \pm \infty$
but $|t|/|r|\leq 1-\epsilon$, $\epsilon>0$ may be an arbitrary small
number (see (\ref{c11})), or\\
2) when $t\ra \pm \infty$ and $r$ is finite or also $r\ra \pm \infty$ but
$|r|/|t| \leq 1-\epsilon $ (eq.~(\ref{c12})).
 We see that the fields themselves are in general infinite at infinity.
 However their first derivatives are finite at infinity and thus
 everywhere; moreover, as the moduli are bounded from above, the first
 derivatives are also bounded.
  The second derivatives and $\exp{q_n}$ are localized in space
 and time being exponentially small at infinity.

To summarize, if we consider the condition for finiteness of the
expressions for $\psi_m,\,\vf,\, F$ we may insert in (\ref{c11}),
(\ref{c12}) only the divergent term,
\be
\label{c13}
    -q_n^{\infty}/2 = \ln |X_n^{\infty}
    |\,=\, \lambda_n\,|\tau|, \quad{\rm for} \,\, |\tau|\ra \infty
\ee
where $\tau =r \ra \pm\infty$ (and $t$ is finite), or $\tau=t\ra\pm \infty$
(and $r$ is finite). The cases when $|r|\ra \infty$ and $|t|\ra (1-\epsilon)r$
or $|t|\ra \infty$ and $|r|=|t|(1-\epsilon)$ can be treated similarly.

Thus we see that the divergent parts of the fields (\ref{c8c}),
{\ref{c8b}), (\ref{c8a}) and of the curvature (\ref{c9}) are
\bean
&& \psi_m^{\infty} \,=\,-\,2\sum_{n=1}^N a_{mn} \gamma_n \lambda_n \,|\tau|,
\quad 3\leq m\leq N\,,
\label{c15a} \\
&& \vf^{\infty} \, =4\sum_{n=1}^N\,\gamma_n\lambda_n\,|\tau|\, ,
\label{c15b}  \\
&& F^{\infty}\, = 4\sum_{n=1}^N\, a_n \gamma_n \lambda_n \,|\tau|.
\label{c15c} \\
&& R^{\infty} \, = -\exp \bigg[ -2|\tau|
 \bigg( 2 \sum_{n=1}^N\, a_n \gamma_n \lambda_n -
 \lambda_{\rm min} \bigg) \bigg] \, ,
\label{c15d}
\eean
where $\lambda_{\rm min}$ is the minimum value of $\lambda_n$ for
 $1 \leq n \leq N$.
 In order to have finite fields $\psi_m,\,\,\vf, \,\,\,F $, the
expressions $\psi_m^{\infty}$, $\vf^{\infty}$  and $ F^{\infty}$ must
vanish\footnote{The condition of finiteness of the fields $\psi_n$ is
physically natural but not absolutely necessary. Also, in general, we need
not require the fields $\phi$ and $F$ to be finite at infinity. As the
first derivatives of all the fields are finite everywhere, including
   infinity, the corresponding energy and momentum densities are finite
   everywhere for any choice of the moduli. }.
 $R$ will be $0$ or $\infty$ depending on the sign of the expression in the
round brackets; it will be finite if that expression is zero. For any
given $N-$ Liouville model the parameters $a_{mn}, \, a_n$ and $\gamma_n$
are fixed and we are free to choose only the parameters $\lambda_n$ that
should satisfy the constraints described above. Of course, not all of
these constraints can be satisfied at the same time.

Suppose that we have solved, e.g., the equations
 $\psi_m^{\infty}=0$ and found
 $\lambda_n$.  Recalling that $\mu_n=\lambda_n+\bar \lambda_n$,
$\nu_n=\lambda_n-\bar \lambda_n$, we immediately can see that in order to
satisfy the constraints (\ref{c4}) $\lambda_n$ (and $\bar \lambda_n$)
should be somehow restricted.  The easiest way to find this restriction is
to use the unit vectors (\ref{18}) that are now constant. Recalling that
\be
\label{c17}
 \hat \xi_k = \hat \gamma_k \, (\lambda_k + \bar \lambda_k) \, , \qquad
 \hat \eta_k = \hat \gamma_k  \, (\lambda_k - \bar\lambda_k), \qquad
 k=2,...,N,
\ee
we define the vectors
\be
\label{c18}
 \hat\lambda_k^+ \, = \, \hat \gamma_k \lambda_k =
 \half (\hat \xi_k + \hat \eta_k)\,, \quad
 \hat \lambda_k^- = \hat \gamma_k  \bar\lambda_k =
 \half \, (\hat \xi_k - \hat \eta_k) \,, \qquad k=2,...,N
\ee
which satisfy two conditions (normalization and orthogonality)
\be
\label{c19}
 \hat \lambda _+^2 \, + \,\hat \lambda_-^2\,=\,1, \quad  \hat \lambda_+ \,
 \cdot\,\hat \lambda_-\,=\,0\,.
\ee
It is easy to see that these two conditions are equivalent to the
constraints (\ref{c4}) .

Using these definitions we can construct the solutions as follows. First take
  any $(N-2)$-dimensional vector $\hat \lambda_+$
  with the norm $\leq 1$ and then take any
vector orthogonal to $\hat \lambda _+$ and having the norm
 $\sqrt {1-\hat \lambda_+^2}$. This gives the general solution.
 However, if $\lambda_k$ are restricted by some additional conditions
 (e.g., finiteness of $\psi_n$), satisfying the condition
 $\hat \lambda_+^2 \,\leq 1\,$ is a nontrivial problem.
 Indeed, if we have derived $\lambda_k$ from some equations,
 we should check that
\be
\label{c20}
    \hat \lambda_+^2 \,\equiv\, \sum_{k=2}^N\, \hat \gamma_k^2 \,
\lambda_k^2\,\leq 1\, ,
\ee
and we shall see in a moment that this condition is really restrictive. If
we have checked it, then the rest is as stated above --
 we take as $\hat \lambda_-$ any vector orthogonal to $\hat \lambda_+$
 and having the norm $\sqrt {1-\hat \lambda_+^2}$.
 Then, according to our general construction, the vectors
\be
\label{c21}
\hat\xi\,=\,\hat \lambda_+\,+\, \hat \lambda_-\,, \quad \hat \eta\,= \,
 \hat \lambda_+\,-\, \hat \lambda_-\,
\ee
define the solution satisfying the constraints (\ref{c4}).

Consider first the equations for $\lambda_n$ given by the conditions
$\psi_m^{\infty}=F^{\infty}=\vf^{\infty}=0$. Using the identities
 (\ref{c10})
we find that the conditions $\psi_m^{\infty}=0,$ $F^{\infty}=0$,
$\vf^{\infty}=0$ give respectively the following equations for
 $y_k\, \equiv \, \gamma_k (\lambda_k\,-1\,)$:
\bean
&& \sum_{k=2}^N \, a_{mk} y_k\, = \, 0 \,, \qquad  m=3,...,N\,,
\label{c24a}\\
&& \sum_{k=2}^N \, y_k\, = \,0\,,
\label{c24c} \\
&& \sum_{k=2}^N \, a_k y_k \, = \, {1\over 2} \, .
\label{c24b}
\eean
In general all these equations have no solution, as we have $N$ linear
equations for $N-1$ parameters $y_k$. The homogeneous equations
 (\ref{c24a}), (\ref{c24c}) always allow for the zero solution $y_k=0$,
which gives $\mu_n=\nu_n=1$ (as $\mu_1=\nu_1=1$). This corresponds to the
static solution with two horizons (see \cite{ATF1}, \cite{ATF2}).

Note that eqs.~(\ref{c24a}), (\ref{c24c})  may have nontrivial ($y_{k}
\not= 0$) solutions if the determinant of the matrix of these equations is
zero. Of course, this is possible only for some special systems and we do
not consider this possibility in general. Note only that for $N=3$ (a
single scalar field) this is impossible because for $a_{33}=a_{32}$ both
   $\gamma_2^{-1}$ and $\gamma_3^{-1}$ should vanish
   (see Appendix to \cite{ATF2}).
   The analysis of the case $N=4$ is much more cumbersome. It shows that
   for any linearly independent set of the vectors $A_n$ satisfying the
   conditions (\ref{aa.050}) the system (\ref{c24a}) has only the trivial
   solution $y_k \equiv 0$. We believe that this statement is true for any
   $N$ but, at the moment, cannot rigorously prove it.

The combination of the equations (\ref{c24a}) and (\ref{c24b}) usually has
a solution because the determinant of these equations is normally
   nonvanishing. For $N=3$ it is easy to find this solution and
   to show that, under certain restrictions,  it defines the moduli satisfying the
   conditions (\ref{c20}). This is demonstrated at the end of this
   Section. We do not treat the general case $N>3$. Although it is easy
   to find the solution of the inhomogeneous linear system
   (\ref{c24a}) and (\ref{c24b}), it is difficult to
   analytically derive the general restrictions on $a_n, a_{mn}$ under
   which the solution satisfies (\ref{c20}).

   Thus, let us first consider a somewhat simpler problem
   of finding the solutions of the system (\ref{c24a})
   which satisfy the restriction (\ref{c20}). Suppose (without
   loss of generality) that not all $a_{mN}$ vanish
   and that the square matrix $a_{mk}$ ($2<k<N-1, \, 3<m<N$)
   is nondegenerate. Then we can, in general,
solve the inhomogeneous system,
\be
\label{c25a}
\sum_{k=2}^{N-1} a_{mk} \, y_k \, = -\,a_{mN}\,y_N , \qquad m=3,...,N \,,
\ee
and express $y_k$ ($k=2, ..., N-1$) in terms of a more or less arbitrarily
   chosen  $y_N$ and of the parameters of the system (of course, we can
   in principle choose any $y_k$ instead of $y_N$).
   The corresponding parameters $\lambda_n$ are
\be
\label{c25}
\lambda_1=1, \qquad \lambda_k \,=\,1\,+\,{y_k\over \gamma_k}, \quad
k=2,...,N\,.
\ee
According to (\ref{c20}) some of the $y_k$ should be negative
  (if $y_k < 0$ for all $k$ then obviously $\lambda^2_+ < 1$
  because $\gamma_k > 0$). Moreover, as
\be
\label{c26}
  \hat \lambda^2_+ = {1\over |\gamma_1|} \sum_{k=2}^N\gamma_k\lambda_k^2 =
{1\over |\gamma_1|} \sum_{k=2}^N \gamma_k \,
 \biggl(1+{y_k\over \gamma_k}\biggr)^2=1+{1 \over |\gamma_1|} \sum_{k=2}^N \,
 \biggl({y_k^2\over \gamma_k} +2y_k \biggr)
\ee
the parameters $y_k$ should satisfy the sufficiently stringent inequality
\be
 \label{c27}
\sum_{k=2}^N \biggl({y_k^2 \over \gamma_k} + 2 y_k \biggr)\, < 0 .
\ee

   As follows from the previous considerations, this inequality
   is equivalent to the fundamental constraints (\ref{c4}) and
   is independent of any equations for $y_k$. The moduli
   $\lambda_n$ related to $y_k$ by (\ref{c25}) and defining
   any solution (\ref{c5a}) ($\bar \lambda_n$ are defined by
   (\ref{c18}) (\ref{c19})) should satisfy this inequality.

   Returning to the system (\ref{c25a}), we
   note that applying (\ref{c27}) to its solution
   gives a quadratic inequality for the arbitrary
   parameter $y_N$ but the implicit dependence
   of this simple inequality on the parameters $a_{mn}$ defining the
   model is so complex that it is rather difficult to make
  any general statement on the solution of this
  problem\footnote{It is not difficult to find the solution for any
  pseudo orthogonal system of parameters $a_{mn}$. However, a general
  analytic solution of the equations is hardly possible.  }.

 This can be seen even in the $N = 3$ case.
 The model is defined by three arbitrary parameters:
 $a_{31}, \, a_{32}, \, a_{33}$.
 The other parameters ($a_n,\, \gamma_n$) can be found by using
the pseudo orthogonality conditions. In this case we have only one
equation (\ref{c24a}) which gives
\be
\label{27a}
  y_2=-{a_{33}\over a_{32}} \, y_3.
\ee
Now, if $a_{33}/a_{32}<0$, we choose $y_3<0$ that automatically gives
 $y_2<0$ and therefore $\hat \lambda_+^2 <1$.
 To find the general solution consider the curve in
 the ($\lambda_2, \lambda_3$) plane defined by the condition
 $\hat \lambda_+^2 =1$, i.e.
\be
\label{c28}
  \hat \lambda_+^2 \, \equiv \, \hat \gamma_2^2 \lambda_2^2 \, +
  \, \hat \gamma_3^2 \lambda_3^2 \, = \, 1 .
\ee
   As both $\lambda_2$ and $\lambda_3$ depend only on $y_3$ they are related by
the equation obtained by excluding this dependence\footnote{To obtain this
we should recall the identities (\ref{c10}) and the definition
 $\hat \gamma_k = \sqrt {\gamma_k / |\gamma_1|}$ introduced previously.}
\be
\label{c30}
 \lambda_2 \, a_{32} \, \hat \gamma_2^2 \, + \,
 \lambda_3 \, a_{33} \, \hat \gamma_3^2 \, = \, a_{31} \,.
\ee
The points ($\lambda_2 > 0, \lambda_3 > 0$) that are inside the ellipse
(\ref{c28}) and belong to the straight line (\ref{c30}) give us all
 possible vectors $\hat \lambda_+$ satisfying required conditions.
To find all possible values of $\mu_n$, $\nu_n$ corresponding to this
vector we take  all possible ${\hat \lambda}_-$ orthogonal  to ${\hat
\lambda}_+$ with the norm ${\hat \lambda}_-^2=1-{\hat \lambda}_+^2$ and
thus find $\bar \lambda_n$. The solution obtained for the $N$-Liouville
problem gives asymptotically finite scalar matter fields as we explained
   earlier.

   We mentioned above that it is possible to find a solution of the equations
   (\ref{c24a}), (\ref{c24b}) satisfying the restriction (\ref{c20}), but
   the restriction requires that the parameters $a_{mn}$ obey certain
   conditions. This statement is easy to illustrate by considering
   the simplest nontrivial case $N=3$. Denoting $a_{3i} \equiv \alpha_i$
   and recalling that (see \cite{ATF2} and Appendix)
\bdm
   a_i = \alpha_i (\alpha_j + \alpha_k) - \alpha_j \alpha_k \,,
   \quad  {\rm where}  \quad (ijk) = (123)_{\rm cycli c} \,,
\edm
   it is easy to find the solution
\be
\label{27b}
    y_2 \, = \, {{\alpha_3} \over {8 a_1 (\alpha_2 - \alpha_3)}} \,,
    \,\,\,\,\,\,\,\,
    y_3 \, = \, -{{\alpha_2} \over {8 a_1 (\alpha_2 - \alpha_3)}} \,.
\ee
    Using the expressions for $\gamma_i$ given in \cite{ATF2}
    (see also Appendix),
    one can show that $y_2 / \gamma_2$ and $y_3 / \gamma_3$ are negative
    if $\alpha_2 < \alpha_1 < \alpha_3$, $\alpha_2<0$, $\alpha_3 > 0$
    (or, if $\alpha_3 < \alpha_1 < \alpha_2$, $\alpha_3 <0$,
    $\alpha_2 >0$). It follows that the condition $\lambda_+^2 < 1$
    is fulfilled and we have the solution satisfying all the necessary
    conditions. In the special case $N=3$ discussed in Appendix we have
     $a_1 = -a_2 = -a_3 = a$, $\alpha_1 = 0$, $\alpha_2 = \alpha < 0$,
     $\alpha_3 = -4a/\alpha > 0$ (if $a > 0$). Simple calculations give
     $y_2 / \gamma_2 = y_3 / \gamma_3 = - \half$ and thus
     $\lambda_+^2 = {1 \over 4}$. More general cases can be treated
     similarly.

The algorithm requested to find similar solutions for $N \geq 4$ is
conceptually clear but in practice it is impossible to give general enough
statements for generic parameters $a_{mn}$. Once the parameters are known
we can study the asymptotic properties of the solutions. However, it is
not clear whether it is possible to give a reasonable a priori
classification of them.
 The same can be said about the possible asymptotic behavior of the
curvature $R$. The singular part of the curvature is
\be
\label{31}
 R^{\infty} = \exp \biggl [-2|\tau| \biggl(2 \sum_{k=2}^N a_k y_k +
 {y_{\bar n} \over \gamma_{\bar n}} \biggr ) \biggr ] \, ,
 \ee
 where $\bar n$ is the number of the minimal $\lambda_n$ (note that
 $y_{\bar n}$ should necessarily be negative).
 In the simplest case $N=3$ one can write an explicit expression of
the expression in round brackets in terms of the three free parameters
$a_{31}, a_{32}, a_{33}$ and thus find when $R^{\infty}$ is zero, infinite
or finite; the last condition is possible only when there exists one
relation between the three parameters (in fact, a quadratic equation for
$a_{31}$).

\section{Separation of variables - simple examples}

Now we obtain the solutions with constant moduli avoiding the complete
explicit solution in dimension 1+1. Instead we employ the generalized
separation of variables. Let us first try to separate the variables $r$
and $t$ in the Liouville equations (\ref{002}). If we simply write
$r=u+v$, $t=u-v$ for all $n$, we obviously end up with one of the naive
reductions, for which $q_n = q_n(r)$ or $q_n = q_n(t)$.

However, taking into account our experience with the solutions of the
$N$-Liouville model, we may try to use a less naive approach and write
different $r_n$, $t_n$ for different $n$:
\be
\label{p.00}
r_n=\,\mu_n u+\nu_n v, \quad t_n=\mu_n u -\nu_n v \,.
\ee
Denoting by `prime' and `dot' the differentiations in $r_n$ and $t_n$ we
obtain
\be
q_n^{''} \, - \, \ddot q_n \,=\, \tilde G_n  \, \exp \, q_n \, , \qquad
 \tilde G_n \, \equiv \, {\tilde g_n \over \mu_n \nu_n}\,.
\label{p.01}
\ee
Now we separate the variables $r_n$ and $t_n$ by writing
\be
\label{p.02}
q_n= \, \xi_n(t_n) + \eta_n(r_n)\,
\ee
   and then differentiate the resulting equation in  $r_n$ or $t_n$. Thus we
immediately find that the following two separations are possible
\bean
\label{p.03}
&& \eta_n^{''} (r_n) \, - C_n \exp \, \eta_n (r_n) \, = \, 0 \,, \qquad
C_n \, \equiv \, \tilde G_n \exp \, \xi_n = {\rm const} \, , \\
\label{p.04}
&& \ddot \xi_n (t_n) \, - \, \bar C_n \exp \, \xi_n (t_n) \, = \, 0 \,, \,
 \qquad
  \bar C_n \, \equiv \, \tilde G_n \exp \, \eta_n = {\rm const} \, .
\eean
This separation gives only half of the solutions with constant moduli in
the $N$-Liouville model. But now one can guess that the separation
procedure will give different results if we apply it to $X_n$ instead of
$q_n$.

Indeed, let us try to separate the same variables (\ref{p.00}) in
equation (\ref{01}) by using the same Ansatz (\ref{p.02}) for $X_n$:
\be
\label{p.06}
X_n(u,v) \, = \, \xi_n(t_n) +\eta_n(r_n)\,.
\ee
Differentiating the resulting equation
\be
\biggl(\xi_n(t_n)+\eta_n(r_n) \biggr)\, \biggl(\eta_n^{''}(r_n) - \ddot
\xi_n(t_n) \,\biggr) +  \eta_n^{'2}(r_n) - {\dot \xi_n}^2(t_n) \,
 =- \half \tilde G_n
\label{p.07}
\ee
in $r_n$ and then in $t_n$, we find that
\be
\eta_n^{'}(r_n) \, \xi_n^{\cdot \cdot \cdot} (t_n) \,- \,
 \eta_n^{'''}(r_n) \,  \dot \xi_n (t_n) = 0 ,
\label{p.07a}
\ee
and therefore $\xi_n$ and $\eta_n$ satisfy the equations
\bean
\label{p.08}
&& \ddot \xi_n (t_n) \,-\,C_n \xi_n (t_n) \,=\, A_n, \\
\label{p.09}
&&\eta_n^{''}(r_n) \,-\,C_n \,\eta_n (r_n) \,=\, B_n\,.
\eean
where $A_n$, $B_n$ and $C_n$ are arbitrary constants. So we obtain
   the solutions,
\bean
\label{p.10}
&&  \xi_n (t_n) = -A_n / C_n  + \, \tilde C_n^-\,{\rm ch}\biggl(\sqrt{C_n}\,
(t_n+\delta_n^{-}) \biggr)\,, \\
\label{p.11}
&& \eta_n (r_n) = -B_n / C_n  + \, \tilde C_n^+\, {\rm ch}\biggl(\sqrt{C_n}\,
(r_n+\delta_n^+) \biggr),
\eean
   with arbitrary integration constants $\tilde C_n^{\pm}$,
   $\delta_n^{\pm}$.
Substituting these functions into eqs.~(\ref{p.07}),  we find that these
  arbitrary constants must
satisfy the equations
\be
C_n\,\biggl(\, (\tilde C_n^+)^2 \,-\, (\tilde C_n^-)^2 \,\biggr) \, = \,
 -{\tilde g_n \over 2\mu_n\nu_n}\,, \qquad A_n\,+\,B_n\,=\,0\,.
\label{p.12}
\ee
We thus see that $X_n$ given by eqs.~(\ref{p.06}), (\ref{p.10}) -
(\ref{p.12}) coincides with (\ref{c5}) - (\ref{c6}) if we take $C_n=1$
 and $\tilde C_n^{\pm}=C_n^{\pm}/\sqrt{\mu_n\nu_n}$
   (or, $C_n = 1/(\mu_n\nu_n)$ and $\tilde C_n^{\pm} = C_n^{\pm}$).  \\

In these considerations we did not use the integrability of the
$N$-Liouville model and one may expect that our approach can be applied to
other nonlinear systems, not necessarily integrable. Indeed, in this way
we treated spherically symmetric static states and cosmologies that are
described by generally nonintegrable equations (see \cite{ATF3}). We thus
expect that solutions similar to those with constant moduli in the
integrable $N$-Liouville theory  may exist also in nonintegrable realistic
(1+1)- dimensional theories. Presumably they can be derived by using some
sort of generalized separation of variables. A serious analysis of this
subject will be attempted elsewhere while here we only give a very simple,
almost trivial example.

Recalling the discussion of `realistic' potentials in Section~2 let us
consider a single scalar field coupled to the dilaton gravity with the
potentials
\be
\label{q.01}
V\,=\,g\,\varphi^{1+a} \,e^{\lambda \psi}, \quad Z=-\varphi\,.
\ee
Introducing the notation $\varphi\equiv e^{\phi}$ we write the main
equations of motion in the form \footnote{As always, we use the `Weyl
frame'  and the light-cone coordinates for the Lagrangian (\ref{aa.010}).}
\bean \label{q.02}
&& \partial_u \partial_v \phi \,+\, \partial_u\phi\,\partial_v\phi\,
+\, \varepsilon g \,e^{F+a\phi+\lambda \psi}\,=\,0, \\
\label{q.03}
&& 2 \partial_u\partial_v \psi+\partial_u\phi \partial_v\psi
+\partial_v\phi \partial_u \psi -\varepsilon g \lambda e^{F+a\phi
+\lambda \psi}=0, \\
\label{q.04}
&& \partial_i^2 \phi +(\partial_i \phi)^2 -\partial_i F \partial_i \phi
+(\partial_i \psi)^2 =0, \quad i=u,v .
\eean

Separations of variables are most convenient if we have bilinear
equations for the unknown functions. In our present example
   eqs.~(\ref{q.04}) are bilinear and we can write one more bilinear equation
combining (\ref{q.02}) with (\ref{q.03}). Defining $\Psi=\psi+\lambda
\phi/2$, we have
\be
\label{q.05}
2\partial_u\partial_v \Psi +\partial_u\Psi \partial_v \phi +\partial_v
\Psi \partial_u \phi = 0.
\ee
The only equation containing the exponential is now eq.~(\ref{q.02}). We
shall not attempt to give a complete study of these equations but simply
write its very special solution
\bean
 \label{q.06}
&& \phi = \alpha \,u +\beta \, v, \qquad \,\,\,\,
 \Psi=\mu\,(\alpha \,u - \beta\, v), \\
 \label{q.07}
&& F = C_0 - \alpha \,u \bigl(a - {\lambda^2 \over 2} + \mu \lambda \bigr)
 - \beta v \, \bigl(a - {\lambda^2 \over 2} - \mu \lambda \bigr) \, ,
\eean
 where $\alpha$, $\beta$ are arbitrary parameters, $C_0={\rm ln} (-\alpha
\beta/g)$ and $\mu^2\equiv \lambda^2/4 \,-1-a$.

Defining $r\equiv u+v$, $t\equiv u-v$, we can rewrite these solutions in
the wave-like (or, pulse-like) form
\bean
\label{q.08}
&& \phi={\alpha+\beta\over 2}\biggl(r+V_1t \biggr), \\
\label{q.09}
&& \Psi={1\over 2} \biggl[(\mu-{\lambda\over 2} )\alpha -(\mu+{\lambda
\over 2}) \beta \biggr]\,(r+V_2t), \\
\label{q.10}
&& F= -{1\over 2} \biggl[ \alpha(a-{\lambda^2\over 2}+\lambda \mu)
+\beta (a-{\lambda^2 \over 2} -\lambda \mu)\biggr] \,(r+V_3t)
\eean
where the velocities of the pulses are
\bean
\label{q.11}
&&V_1= {\alpha-\beta\over 2}, \\
\label{q.12} && V_2={(\mu -{\lambda\over 2} )\alpha
+(\mu+{\lambda\over 2})\beta \over (\mu-{\lambda\over 2} )\alpha
-(\mu+{\lambda\over 2}) \beta}, \\
\label{q.12a} && V_3={\alpha(a-{\lambda^2\over 2}+\mu\lambda)-\beta
   (a -{\lambda^2\over 2} -\mu\lambda) \over \alpha(a-{\lambda^2\over 2}
+\mu\lambda)+\beta (a-{\lambda^2\over2} -\mu\lambda)}\,.
\eean
 Choosing the gauge $\alpha=\beta$ (`static' $\phi$ ) we have
\be
\label{q.13}
V_1=0, \quad V_2=-{2\mu\over \lambda}, \quad V_3=-{2\mu \over \lambda}
\biggl( 1-{2a\over \lambda^2}\biggr)^{-1}\,.
\ee

The model in which we have found this solution belongs to the class of
two-dimensional theories obtained by standard dimensional reductions from
higher-dimensional theories. We thus see that  the wave - like solutions
can be met not only in the integrable models with $N=1$ but also in quite
realistic and apparently nonintegrable theories. So one may hope that the
connection between static, cosmological and wave - like solutions is a
general feature of the theory of gravity \footnote{In \cite{ATF3} we
demonstrated that the duality relation between static states and cosmology
is present in the case of spherically reduced gravity coupled to scalar
matter fields.}.

\section{Conclusion and outlook}

The concrete results of this paper are as follows. We have generalized the
analytic expression for the solution of the equations and constraints of
  the (1+1)-dimensional $N$-Liouville  theory, (\ref{04}), (\ref{05}). We
  have introduced a very convenient gauge (\ref{17}), in which the main
physical properties of the solutions are particularly transparent, and
then derived the representation of the solution in terms of moduli
 functions $\hat \xi(u)\, \in \, S^{(N-2)}$,
 ~$\hat \eta(v)\, \in \, S^{(N-2)}$.
    The reduction of the moduli space  to the space  of points
(${\hat \xi}^{(0)}$, ${\hat \eta}^{(0)}$), belonging to $S^{(N-2)}$
introduces an interesting new class of solutions  of the $N$-Liouville
theory, ({\ref{c2}), (\ref{c3} ). When ${\hat \xi}^{(0)}$ =${\hat
\eta}^{(0)}$, these solutions  can be interpreted as either static states
   or as cosmologies. When ${\hat \xi}^{(0)} \neq {\hat \eta}^{(0)}$ the
   corresponding solutions depend on both variables $r$ and $t$.

In general, the solutions have singularities at finite space-time points
   $(t,\,\,r)$ and at infinity. However there exists a subclass  of solutions
having no singularities at finite points, (\ref{c5}), (\ref{c6}). A
solution of this (nonsingular) subclass may be usually infinite at the
space and/or time infinity, but we showed that some of them are finite
both for $r\rightarrow \pm\infty$ and for $t \rightarrow \pm\infty$. For
the solutions to be finite at the space and time boundaries we have
   formulated explicit constraints that are simple to solve; however, they
   cannot be solved in general and so we cannot derive a general formula
   for the solution of these constraints in an arbitrary $N$-Liouville model.
   We therefore constructed an algorithm for solving this problem and
illustrated it by a very simple example for $N=3$.

   The solutions without singularities at finite points
   have finite densities of energy and momentum and
the shape of their first derivatives is step - like, similar to some
topological solitons. So, they resemble solitons but their energy is not
   localized in space and for this reason we cannot call them solitons.
   The curvature $R$ for these solutions is finite at finite points
   and may diverge at
infinity (this depends on all the parameters defining the solution). Thus,
   in spite of their apparent simplicity,
   the solutions defined by constant moduli are rather complex objects
strongly interconnected to the geometry of the space-time.

We believe that very similar waves can be obtained in realistic,
  nonintegrable theories. There they should be looked for with the aid of
the generalized separation of variables. We demonstrated here that the
wave solutions of the integrable (1+1)-dimensional theory can be obtained
without any knowledge of its integrability; we also gave an example of the
simplest solutions in a nonintegrable theory.

   In future, we plan to concentrate on the derivation of waves, similar to
   those found here, in realistic nonintegrable theories.
     Simple enough and interesting waves that should be studied from the point
     of view advocated here are cylindrical waves \cite{Einstein} and plane
     waves \cite{Szekeres}.
We also hope to find a closer connection between static states,
cosmologies and waves in completely realistic theories.
     Some physical connections between waves and black holes were
     discussed in literature (see, e.g. \cite{Arefeva} and references therein).
     A possible role of gravitational waves in cosmology was also studied
     in some detail (see, e.g. \cite{Feinstein} and \cite{Bozza}).
     At first sight, our model results may look not so directly related to these
     investigations but being conceptually and technically very simple they
     can help to find a new systematic approach to a deeper understanding of
     these complex phenomena.

\section{Appendix}

    Here we give parameters of a few typical $N$-Liouville models, some of which can
    be obtained by dimensional reduction of the higher-dimensional supergravity
    theories. We first construct the general $N=4$ model. Using the
    formulas of Appendix to \cite{ATF2} we can immediately write the
    general expressions for $a_i$ and $\gamma_i$ ($i = 1,2,3$) in the
    general $N=4$ case. Introducing the notation $a_{3n} \equiv \alpha_n$,
    $a_{n} \equiv \beta_n$ ($n=1,...,4$) we have
\bean
&& 4a_i =  \alpha_i (\alpha_j + \alpha_k) - \alpha_j \alpha_k \,+\,
 \beta_i (\beta_j + \beta_k) - \beta_j \beta_k \,,
\label{A1} \\
&& \gamma_i = (\alpha_i - \alpha_j) (\alpha_i - \alpha_k) \,+\,
(\beta_i - \beta_j) (\beta_i - \beta_k) \,,
\label{A2}
\eean
   where $\alpha_i$, $\beta_i$ are arbitrary parameters and
   $(ijk) = (123)_{\scriptsize \textrm{cyclic}}$. The unknown
   parameters $a_4$, $\alpha_4$, $\beta_4$ can be derived by applying the
   general procedure of ref.~\cite{ATF2}. In the $N=4$ case it gives three
   inhomogeneous linear equations for these three parameters. The solution
   is
\bean
&& a_4 = -{1 \over \Delta} \bigl[a_1 (\alpha_2 \beta_3 - \alpha_3 \beta_2) \,+\,
            a_2 (\alpha_3 \beta_1 - \alpha_1 \beta_3) \,+\,
            a_3 (\alpha_1 \beta_2 - \alpha_2 \beta_1) \bigr] \,,
\label{A3}\\
&& \alpha_4 = +{2 \over \Delta}
      \bigl[a_1 (\beta_2 - \beta_3) \,+\,
            a_2 (\beta_3 - \beta_1) \,+\,
            a_3 (\beta_1 - \beta_2) \bigr] \,,
\label{A4}\\
&& \beta_4 = -{2 \over \Delta}
      \bigl[a_1 (\alpha_2 - \alpha_3) \,+\,
            a_2 (\alpha_3 - \alpha_1) \,+\,
            a_3 (\alpha_1 - \alpha_2) \bigr] \,,
\label{A5}\\
&& \Delta \,=\, \bigl[\alpha_1 (\beta_2 - \beta_3) \,+\,
            \alpha_2 (\beta_3 - \beta_1) \,+\,
            \alpha_3 (\beta_1 - \beta_2) \bigr] \,.
\eean
  Alternatively, we can use  (\ref{A1}), (\ref{A2})  and the `sum rules'
  (\ref{c10}) to derive $\gamma_4$, $a_4$, $\alpha_4$, $\beta_4$ in terms
  of the previously found $a_i$, $\gamma_i$ ($i = 1,2,3$).

  For $N=5$, deriving the  parameters of the models becomes much
  more cumbersome. Moreover, the general expressions become useless for
  practical analytical computations, even in the simple problems treated
  in the main body of our paper. Fortunately enough, realistic
  $N$-Liouville theories obtained from higher-dimensional theories
  often have much simpler structure than in the general case. We mentioned
  above the important examples:
\be
   \textrm{I}. \,\,\, -a_1 = a_2 = ... = a_N = -a \,; \quad
   \textrm{II}. \,\,\, -a_1 = -a_2 = a_3 = ... = a_N = -a \,.
\label{A6}
\ee
   To find the expressions for the parameters $a_{mn}$ of these models it
   is convenient to use simple geometrical considerations instead of
   applying the general procedure. With this aim, we write the vectors
   $A_n$ introduced in Section~2 in the form
\be
 A_n \equiv (1 + a_n \,, 1 - a_n\,, \overrightarrow A_n) \,, \quad
 \overrightarrow A_n \equiv (a_{3n}\,,..., a_{Nn}) \,,
\label{A7}
\ee
  where $\overrightarrow A_n$ are euclidean $(N-2)$-vectors that should
  satisfy the restrictions following from eqs.~(\ref{aa.050}).
  For the models (\ref{A6}), one can
  explicitly solve this problem for arbitrary $N$ but here we only write
  the result for $N=4$. For the type I models:
\bean
&&  {\overrightarrow A}_1 \,=\, (0,0), \,\,\, \gamma_1^{-1} = -4a ; \quad
  {\overrightarrow A}_2 \,=\, (\alpha ,0) , \,\,\,
  \gamma_2^{-1} = \alpha^2 + 4a ;
  \nonumber \\
&&  {\overrightarrow A}_3 \,=\, (-4a / \alpha , \,\beta); \quad \quad \quad
  \gamma_3^{-1} = (\alpha^2 \beta^2 + 4a \alpha^2 +
  16a^2) / \alpha^2 \,;
  \nonumber \\
&&  {\overrightarrow A}_4 \,=\,
  (-4 / \alpha , -4a (\alpha^2 + 4a)/ \alpha^2 \beta); \quad
  \gamma_4 \,=\, -(\gamma_1 + \gamma_2 + \gamma_3) .
\label{A8}
\eean
For the type II models:
\bean
& {\overrightarrow A}_1 \,=\, (\alpha,0), \,\,\,
  \gamma_1^{-1} = \alpha^2 - 4a ; \quad
  {\overrightarrow A}_2 \,=\, (4a/\alpha \,,0) , \,\,\,\,\,\,
  \gamma_2^{-1} = 4a(4a - \alpha^2)/\alpha^2 ;
  \nonumber \\
&  {\overrightarrow A}_3 \,=\, (0, \beta); \,\,\,
  \gamma_3^{-1} = \beta^2 + 4a \,; \quad
  {\overrightarrow A}_4 \,=\, (0, -4a / \beta), \,\,\,
  \gamma_4^{-1} = 4a (\beta^2 + 4a)/ \beta^2 .
\label{A9}
\eean
 Using these results the reader can easily derive the vectors
 ${\overrightarrow A}_n$ for the type I and type II $N=5$ models
 and generalize these formulas to arbitrary $N$.

 The examples (\ref{A6}) are interesting not only because of their
 simplicity. As we
 mentioned in Section~2, they are generic in the context of dimensional
 reduction of some higher-dimensional supergravity theories (see, e.g. the model
 of ref.~\cite{Kiem} the parameters of which are given in (\ref{110b})).

\section{Acknowledgments} One of the authors (A.T.F.) very much appreciates
 the support of the Department of Theoretical Physics of the University of Turin
 and of INFN (Section of Turin).

 This work was supported in part by the Russian Foundation for Basic
 Research (Grant No. 06-01-00627-a).

\end{document}